\documentclass[runningheads]{llncs}
\usepackage{graphicx}
\usepackage{float}
\usepackage{booktabs} % for professional tables
\usepackage{amsmath,amssymb,amsfonts}
\newcommand{\samelineand}{\quad\quad}
\usepackage[misc]{ifsym}
\usepackage{hyperref}

\begin{document}
\title{CT-SGAN: Computed Tomography \\Synthesis GAN\thanks{Supported by Imagia Inc.\\ Deep Generative Models workshop @ MICCAI 2021 (Peer-reviewed)}}
\vspace{-1ex}
%
%\titlerunning{Abbreviated paper title}
% If the paper title is too long for the running head, you can set
% an abbreviated paper title here
%
\author
{%
    Ahmad Pesaranghader\inst{1, 2, 3\, }\textsuperscript{\Letter}
    \and
    Yiping Wang\inst{1, 4}
    \and
    Mohammad Havaei\inst{1}
}

\institute
{
    \samelineand \inst{1} Imagia Inc. 
    \samelineand \inst{2} Quebec AI Institute (Mila)\\
    \samelineand \textsuperscript{ \Letter} \href{mailto:pesarana@mila.quebec}{pesarana@mila.quebec}\\
    \samelineand \inst{3} McGill University
    \samelineand \inst{4} University of Waterloo
    \vspace{-1.5ex}
}

\authorrunning{A. Pesaranghader et al.}

% First names are abbreviated in the running head.
% If there are more than two authors, 'et al.' is used.
%
%\institute{Anonymous Institute}
%
\maketitle              % typeset the header of the contribution
\begin{abstract}
Diversity in data is critical for the successful training of deep learning models. Leveraged by a recurrent generative adversarial network, we propose the CT-SGAN model that generates large-scale 3D synthetic CT-scan volumes ($\geq 224\times224\times224$) when trained on a small dataset of chest CT-scans. CT-SGAN offers an attractive solution to two major challenges facing machine learning in medical imaging: a small number of given i.i.d. training data, and the restrictions around the sharing of patient data preventing to rapidly obtain larger and more diverse datasets. We evaluate the fidelity of the generated images qualitatively and quantitatively using various metrics including Fréchet Inception Distance and Inception Score. We further show that CT-SGAN can significantly improve lung nodule detection accuracy by pre-training a classifier on a vast amount of synthetic data. 

\end{abstract}
\section{Introduction}
Recently, deep learning has achieved significant success in several applications including computer vision, natural language processing, and reinforcement learning \cite{henderson2018deep}\cite{jiang2017trajectorynet}\cite{lao2019dual}\cite{pesaranghader2021imputecovnet}. However, large amounts of training samples, which sufficiently cover the population diversity, are often necessary to develop high-accuracy machine learning and deep learning models \cite{pesaranghader2018one}\cite{pesaranghader2019deepbiowsd}. Unfortunately, data availability in the medical image domain is quite limited due to several reasons such as significant image acquisition costs, protections on sensitive patient information, limited numbers of disease cases, difficulties in data labeling, and variations in locations, scales, and appearances of abnormalities. Despite the efforts made towards constructing large medical image datasets, options are limited beyond using simple automatic methods, huge amounts of radiologist labor, or mining from radiologist reports \cite{jin2018ctrealistic}. Therefore, it is still a challenge to generate effective and sufficient medical data samples with no or limited involvement of experts.

The production of synthetic training samples is one tempting alternative. However, in practice, it is less appealing due to weaknesses in reproducing high-fidelity data \cite{jin2018ctrealistic}. The advancement in generative models such as the generative adversarial networks (GANs) \cite{goodfellow2014generative}, however, is creating an opportunity in producing real-looking additional (training) data. This possibility has been enhanced with refinements on fully convolutional \cite{radford2015unsupervised} and conditional GANs \cite{mirza2014conditional}. For example, Isola et al. extend the conditional GAN (cGAN) concept to predict pixels from known pixels \cite{isola2017image}. Within medical imaging, Nie et al. use a GAN to simulate CT slices from MRI data \cite{nie2017medical}. For lung nodules, Chuquicusma et al. train a simple GAN to generate simulated images from random noise vectors, but do not condition based on surrounding context \cite{chuquicusma2018fool}. Despite recent efforts to generate large-scale CT-cans with the help of GANs \cite{sun2020hierarchical}\cite{park2021realistic}, to the best of our knowledge, all of these generative models have one or more important pieces missing which are: (1) the (non-progressive) generation of the whole large-scale 3D volumes from scratch with a small number of training samples, (2) the generation of CT-scans without working with a sub-region of a volume or translating from one domain/modality to another domain/modality, and (3) examining their generative model with real-life medical imaging problems such as nodule detection. The absence of these missing pieces can be due to large amounts of GPU memory needed to deal with 3D convolutions/deconvolutions \cite{shin2018medical}\cite{van2019chest}\cite{mirsky2019ct}. This limitation makes even the most well-known GANs for the generation of high-resolution images \cite{karras2017progressive}\cite{zhang2017stackgan} impractical once they are applied to the 3D volumes. On the other hand, the generation of large-scale CT-scan volumes is of significant importance as in these scans, fine parenchymal details such as small airway walls, vasculature, and lesion texture, would be better visible which in turn lead to more accurate prediction models.

\begin{figure}[H]
	\centerline{\includegraphics[scale=0.68]{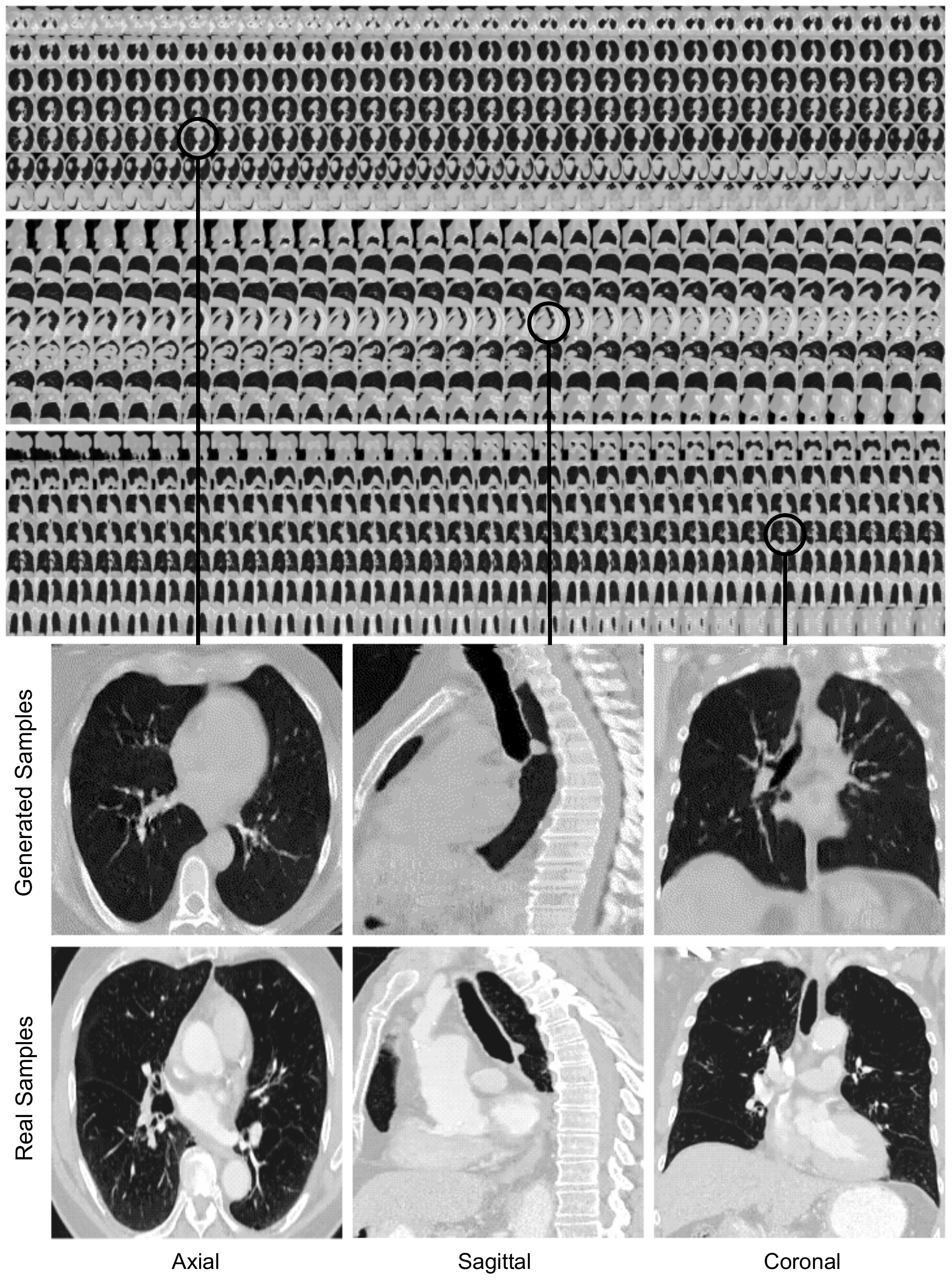}}
	\caption{Generated samples from the CT-SGAN model ($224\times224\times224$) compared with samples of the same positions from a real CT-scan. The upper part of the figure also demonstrates CT-SGAN, with the help of its BiLSTM network as well as slab discriminator, is capable to learn anatomical consistency across slices within every 900 real training CT-scan volumes reflecting the same behavior in all three perspectives of axial, sagittal, and coronal at generation time. See Appendix for more samples.}
	\label{CTSGAN-sample_result}
\end{figure}

In this work, we propose a novel method to generate large-scale 3D synthetic CT-scans ($\geq 224\times224\times224$) by training a recurrent generative adversarial network with the help of a small dataset of 900 real chest CT-scans. As shown in Figure \ref{CTSGAN-sample_result}, we demonstrate the value of a recurrent generative model with which the volumes of CT-scans would be generated gradually through the generation of their sub-component slices and slabs (i.e., series of consecutive slices). By doing so, we can subvert the challenging task of generating large-scale 3D images to one with notably less GPU memory requirement. Our proposed 3D CT-scan generation model, named CT-SGAN, offers a potential solution to two major challenges facing machine learning in medical imaging, namely a small number of i.i.d. training samples, and limited access to patient data. We evaluate the fidelity of generated images qualitatively and quantitatively using Fréchet Inception Distance and Inception Score. 
We further show training on the synthetic images generated by CT-SGAN significantly improves a downstream lung nodule detection task across various nodule sizes. Hence, our contributions are twofold: 
\vspace{-3.5ex}
\begin{itemize}
    \item We propose a generative model capable of synthesizing 3D images of lung CT. The proposed CT-SGAN leverages recurrent neural networks to learn slice sequences and thus has a very small memory footprint.
    \item We demonstrate a successful use case of pre-training a deep learning model on 3D synthetic CT data to improve lung nodule detection.  
\end{itemize}

\section{Methods}

\begin{figure*}[ht]
	\centerline{\includegraphics[scale=0.59]{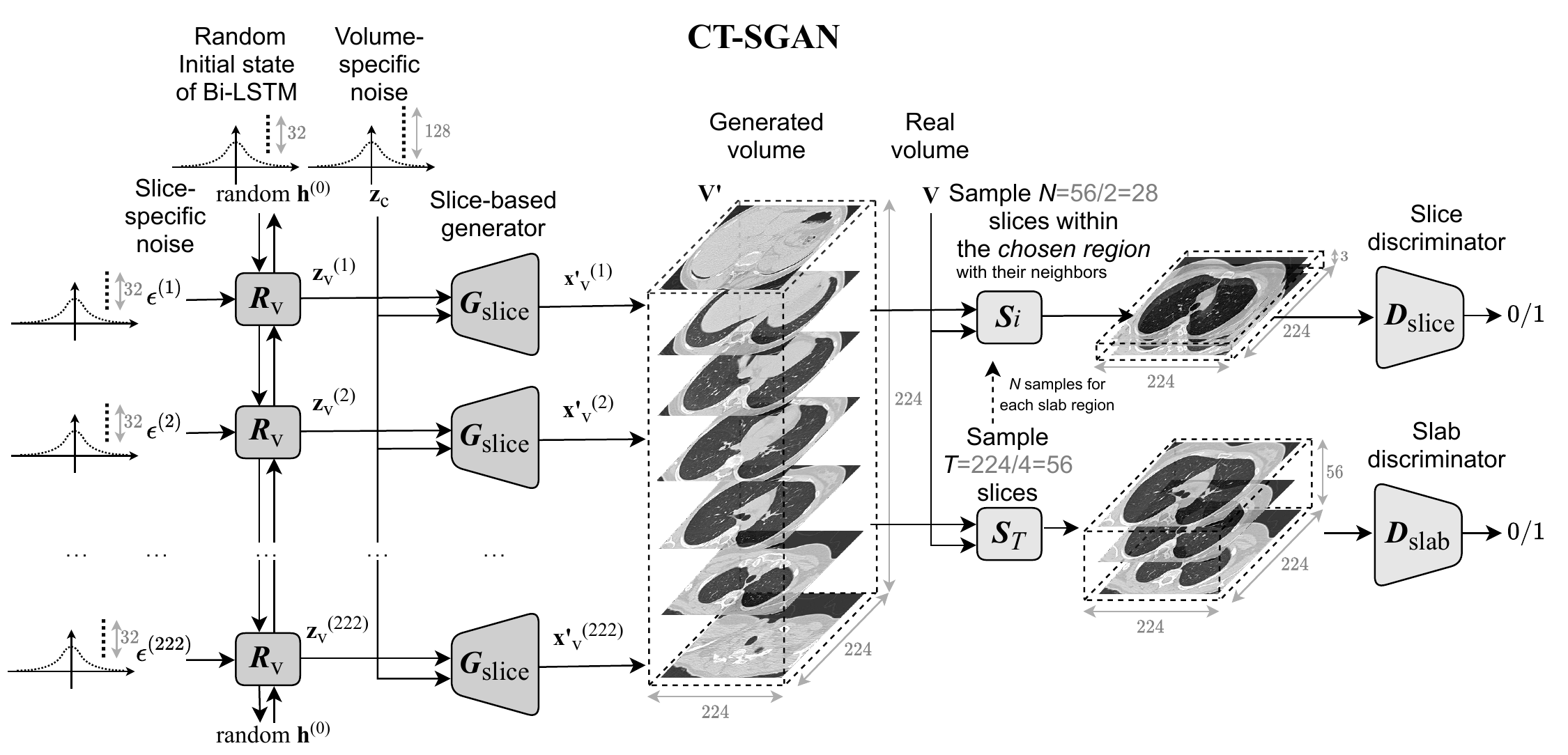}}
	\caption{CT-SGAN has 4 sub-networks: the recurrent neural network (BiLSTM), $R_v$; the slice generator, $G_{slice}$; the slice and the slab discriminators, $D_{slice}$ and $D_{slab}$. $G_{slice}$ generates a volume by getting $z$s from $R_v$ concatenated with the constant noise. The discriminators verify the consistency and the quality of the generated slices with respect to original CT-scan volumes.}
	\label{CTSGAN-network}
\end{figure*}

The structure of the CT-SGAN model is depicted in Figure \ref{CTSGAN-network}. The model architecture is closely related to a work done by Tulyakov et al. \cite{tulyakov2018mocogan} for the generation of videos. In our work, however, instead of considering frames of a video that are meant to be generated sequentially in terms of time, we have a series of slices (i.e., slabs from a volume) that are needed to be generated consecutively in terms of space. Some major modifications in the network designs such as the critical change in $z$ size, consideration of gradient penalty, sampling from the slab regions, consideration of Bidirectional Long/Short-Term Memory (BiLSTM) network for $R_v$ instead of Gated Recurrent Units, and employment of 3D slices differentiate CT-SGAN from their study. These changes specifically help the generation of large-scale 3D CT-scans to be efficient and with high fidelity to real data.

Succinctly, CT-SGAN generates a CT-scan volume by sequentially generating a single slice each of size $224\times224~(\times3)$; from these 3D slices, which lead to more fine-grain quality, the collection of center slices form a volume when piled on top of each other (the first and the last 1D slices in the volumes are the neighbors of their associate 3D slices; hence $222$ 3D slices in total). At each time step, a slice generative network, $G_{slice}$, maps a random vector $z$ to a slice. The random vector consists of two parts where the first is sampled from a constant patient-specific subspace and the second is sampled from a variable slice-specific subspace. Patient-specific noise can also cover the overall color range and contrast within a volume (e.g., scanner-specific information). Since contents of CT-scans slices in one volume usually remains the same, we model the patient/scanner-specific space using a constant Gaussian distribution over all slices for a particular volume. On the other hand, sampling from the slice-specific space is achieved through a BiLSTM network ($R_v$) where the network is fed with random inputs ($\epsilon_i$ and $h_0$), and its parameters are learned during training. Despite lacking supervision regarding the decomposition of general (patient/scanner-specific) content and slice-specific properties in CT-scan volumes, we noticed that CT-SGAN can learn to disentangle these two factors through an adversarial training scheme as the absence of either noise hurt the overall quality. Both discriminators $D_{slice}$ and $D_{slab}$ play the judge role, providing criticisms to $G_{slice}$ and $R_v$. The slice discriminator $D_{slice}$ is specialized in criticizing $G_{slice}$ based on individual CT-scan slices. It is trained to determine if a single slice is sampled from a real CT-scan volume, $v$, or from $v^{\prime}$ with respect to the slice position. On the other hand, $D_{slab}$ provides criticisms to $G_{slice}$ based on the generated slab. $D_{slab}$ takes a fixed-length slab size, say T (center) slices, and decides if a slab was sampled from a real CT-scan volume or from $v^{\prime}$. The adversarial learning problem writes as:
\begin{equation}
\max _{G_{\mathrm{slice}}, R_{\mathrm{v}}} \min _{D_{\mathrm{slice}}, D_{\mathrm{slab}}} \mathcal{F}_{\mathrm{volume}}\left(D_{\mathrm{slice}}, D_{\mathrm{slab}}, G_{\mathrm{slice}}, R_{\mathrm{v}}\right)
\end{equation}
where the vanilla Jensen-Shannon divergence objective function $\mathcal{F}_{\mathrm{volume}}$ is ($N$ is the number of slices sampled from a slab region):
\begin{equation}
\begin{array}{l}
\sum_{i=1}^N\big[{\mathbb{E}_{v}\left[-\log D_{slice}\left(S_{i}(v)\right)\right]}+ 
\\{\mathbb{E}_{v^{\prime}}\left[-\log \left(1-D_{slice}\left(S_{i}({v^{\prime}})\right)\right)\right]\big]+}
\\ {\mathbb{E}_{v}\left[-\log D_{slab}\left(S_{T}(v)\right)\right]}+  
\\ {\mathbb{E}_{v^{\prime}}\left[-\log \left(1-D_{slab}\left(S_{T}({v^{\prime}})\right)\right)\right]}\end{array}
\end{equation}
We train CT-SGAN using the alternating gradient update algorithm \cite{goodfellow2016nips}. Specifically, in one step, we update $D_{slice}$ and $D_{slab}$ while fixing $G_{\mathrm{slice}}$ and $R_{\mathrm{v}}$. In the alternating step, we update $G_{\mathrm{slice}}$ and $R_{\mathrm{v}}$ while fixing $D_{slice}$ and $D_{slab}$.

\section{Datasets and Experimental Design}

The evaluation of the generated CT-scans was designed to be done under two scrutinies: (1) qualitative inspection of the generated volumes from CT-SGAN where the diverse variation and consistency across all three views of axial, coronal, and sagittal were met (2) quantitative demonstration that synthetic data from CT-SGAN are valuable to build deep learning models for which limited training data are available. We evaluate the efficacy of data augmentation by three nodule detection classifier experiments (i) training with only real dataset (ii) training with only 10,000 synthetic volumes (iii) training with 10,000 synthetic volumes as a pretraining step and then continue to train on the real dataset (i.e., fine-tuning). 

\subsection{Dataset preparation}
The dataset preprocessed and prepared for the training and evaluation contained $1200$ volumes of CT-scans; the first half (clear of any nodules) was from the National Lung Screening Trial (NLST) study \cite{national2011national} and the second half (containing nodules) was from the Lung Image Database Consortium (LIDC) reference database \cite{armato2011lung} (i.e., the dataset covers at least two distributional domains), where both are publicly available. This combined dataset was divided into 3 stratified non-overlapping splits of training, validation, and test. Explicitly, of this combined dataset (referred as real dataset hereafter), 900 CT-scans were used for CT-SGAN training (as well as the nodule detection classifier when needed), 150 scans for validation of the nodule detection classifier, and the nodule detection results were reported on the remaining 150 CT-scans. The CT-SGAN model was trained only on the training split; however, since CT-SGAN generates samples unconditioned on the presence of the nodules, the nodules from LIDC were removed in advance (i.e., CT-SGAN generates nodule-free CT-scans). Regarding nodule detection experiment, as the real CT-scans came from two resources (i.e., LIDC scanners and NLST scanners), one of which contained nodules, the training, validation and test dataset of the nodule detection classifier was created by following Figure~\ref{fig:cls}(a) to create an unbiased (source- and device-agnostic) dataset. For this purpose, a \textit{nodule injector} and a \textit{nodule eraser} were adopted based on a lung nodule simulation cGAN \cite{jin2018ctrealistic}.

The nodule injector was trained on the LIDC data, which contains nodules information such as location and radius. At training time the inputs were the masked central nodules regions and the volumes of interest containing the nodules, while at inference mode the input was only the masked central nodules regions, see Appendix. The outputs were the simulated nodules when provided with masked regions. In a similar fashion to lung nodule simulation cGAN \cite{jin2018ctrealistic}, the nodule eraser was trained to remove nodules on nodule-injected NLST samples. At training time, the inputs of the eraser were the volumes of interest with and without a nodule, and the model learned how to replace the nodules with healthy tissues at inference time, see Appendix.

To mitigate the device noise in the real dataset, the LIDC data was divided evenly and the nodule eraser was applied to remove the nodules from half of the LIDC scans. Similarly, the NLST data was split evenly and the nodule injector was applied to insert the nodules into half of the NLST scans.

To obtain the synthetic volumes for augmentation, the nodule eraser was first applied to the LIDC data to ensure the CT-SGAN training set was nodule-free. We argue the synthetic scans were nodule-free as the nodules in training data were removed. The trained nodule injector was employed to randomly insert nodules inside the lungs of the synthetic volumes, and the number of nodules to be inserted was determined by the nodule amount distribution of the LIDC dataset. The 10,000 synthetic data augmentation were created by injecting nodules into half of the volumes and leave the rest untouched.

\section{Results and Discussion}
\subsection{Qualitative evaluation}
The visual qualitative evaluation of the generated volumes was studied based on three criteria: (1) Anatomical consistency of the generated slices, (2) Fidelity of generated slices to the real ones, and (3) diverse generation of CT-scans. Regarding the first two, Figure \ref{CTSGAN-sample_result} shows these requirements were met as in thousand of generated CT-scans we rarely noticed any anomalies. For the high quality of the slices and slabs, we observed consideration of 3D slices and the inclusion of both patient- and slice-specific noises played important roles. As to the diversity in the generated CT-scans, i.e. to avoid mode collapse, and also to ensure stability in training, the discriminators' losses contained the gradient penalty discussed in \cite{mescheder2018training} introducing variations in the generated volumes. While CT-SGAN was preferably trained with Wasserstein loss~\cite{arjovsky2017wasserstein}, we did not notice a drastic change when the vanilla Jensen-Shannon loss was employed. Also, even though artifacts could appear in the generated CT-scans, the presence of them was partially related to the noise in real CT-scans produced by scanners.

\subsection{Quantitative evaluation}

\begin{figure}[ht]
	\centerline{\includegraphics[scale=0.45]{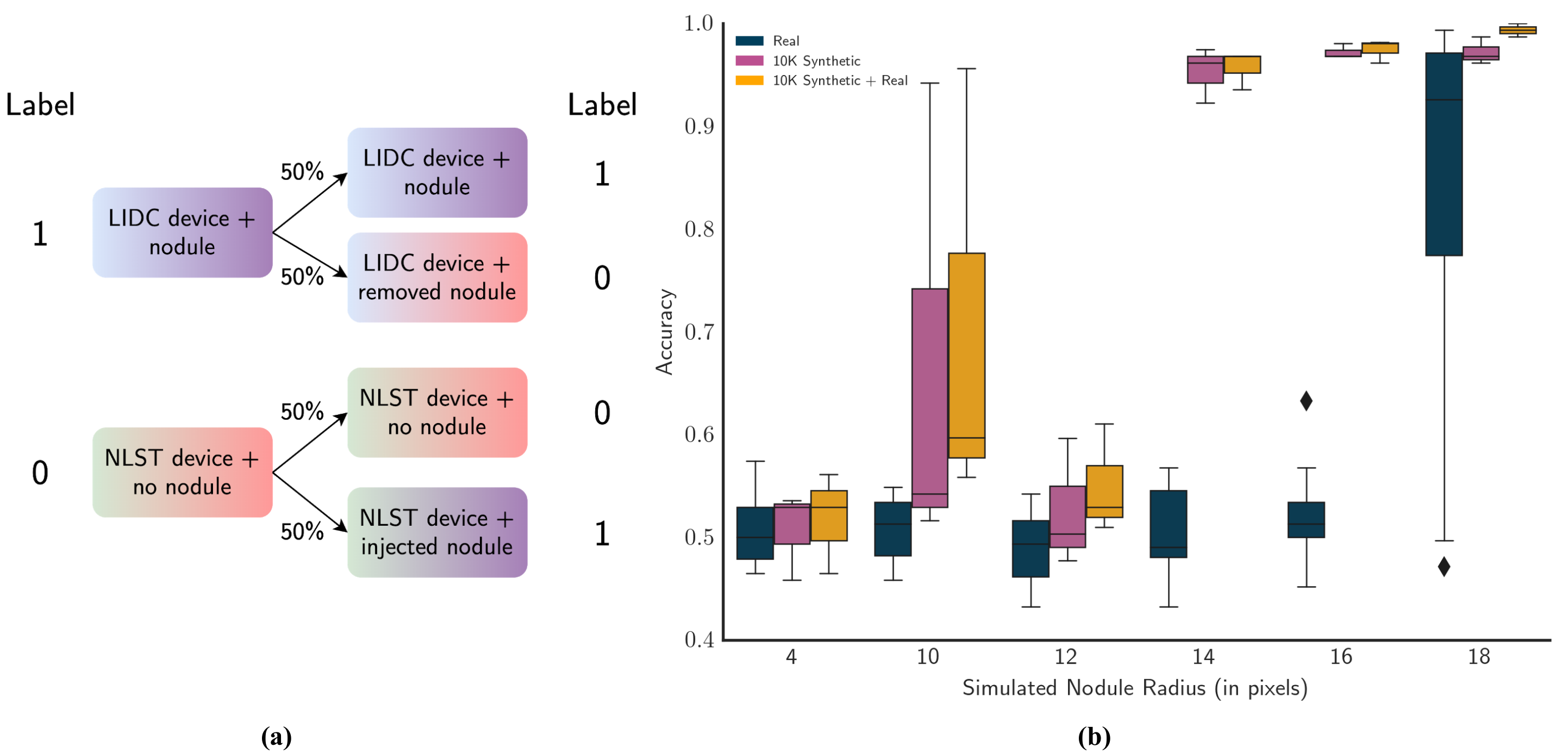}}
	\caption{\textbf{(a)} The real dataset consists of both LIDC and NLST data. To overcome the effect of device noise on the nodule detection classifier, a nodule \textit{injector} and a nodule \textit{eraser} were trained to mix the nodule information and device information to alleviate the device bias. \textbf{(b)} Results for classifiers trained on real and synthetic images. Accuracy is provided for the balanced binary test set. We observe that (pre-)training the classifier on a large amount of synthetic data will notably improve the nodule detection performance. }
	\label{fig:cls}
\end{figure}

3D-SqueezeNet~\cite{iandola2016squeezenet} was used as the nodule detection classifier. Input volumes were normalized between 0 and 1, and the classifier predicts the existence of nodules as a binary label (0  or 1). Adam optimizer~\cite{kingma2017adam} was used with learning rate = 0.0001, $\beta_1$ = 0.9, $\beta_2$ = 0.999. 3 different seeds were used to initialize the classifier and choose the location and number of nodules to inject into the synthetic volumes. Moreover, 6 different sizes of simulated nodules were compared, see Appendix for the distribution of nodule radius in LIDC dataset. Figure~\ref{fig:cls}(b) summarized the nodule detection classification results. We observe that with a larger number of synthetic training data (10,000 generated CT-scans), the trained classifiers have better performance when compared with the classifier trained with 900 real volumes; the accuracy improvement is significant for the nodule sizes of 14 and 16. Also, nodule classification accuracy increases even further by pre-training the classifier on synthetic data and then fine-tuning on the 900 real volumes. 

Fréchet Inception Distance (FID)~\cite{heusel2018gans} and Inception Score (IS)~\cite{salimans2016improved} were also computed, as shown in Table~\ref{tab:inception-score}. FID measures the disparity between the distribution of generated synthetic data and real volumes. IS indicates how diverse a set of generated CT volumes is. To compute FID, we randomly selected two different scans in the data source (real or synthetic) and computed the FID slice by slice in order. Similarly, IS was also computed slice-by-slice. The average IS for the synthetic data is 2.3703 while the average IS for the real data is 2.2129. We believe that the IS for real volumes is lower due to the limited real volumes (only 900 scans). The FID between synthetic data and real data is 145.187. As a point of reference, the FID between two splits of the real data is 130.729. As the generation of small-size volumes (e.g., $128\times128\times128$) with a vanilla 3D GAN, and then resizing the generated scans (at the cost of losing details and diversity) determined the baseline in our model comparison Table~\ref{tab:inception-score} also provides that.

\begin{table}[h!]\centering
\caption{Inception Score and Fréchet Inception Distance for the real scans and the synthetic scans generated by vanilla 3D GAN and CT-SGAN.}
\label{tab:inception-score}
%\scriptsize
\begin{tabular}{lcccc}\toprule
Data Source &IS~($\uparrow$) & &FID~($\downarrow$) \\\midrule
Real Data &2.21 $\pm$ 0.21 & &130.72 $\pm$ 31.05 \\
3D GAN Synthetic Data ($128^3$) &2.16 $\pm$ 0.26 & &206.34 $\pm$ 59.12 \\
CT-SGAN Synthetic Data ($224^3$)~~ &2.37 $\pm$ 0.19 & &145.18 $\pm$ 25.97 \\
\bottomrule
\end{tabular}
\end{table}

\section{Conclusions}
We introduced CT-SGAN, a novel deep learning architecture that can generate authentic-looking CT-scans. We quantitatively demonstrated the value of data augmentation using CT-SGAN for the nodule detection task. By pretraining the nodule detection classifier on a vast amount of synthetic volumes and fine-tuning on the real data, the performance of the classifier improved notably. For future work, we aim to generate a larger size of CT-scans with the proposed model, as well as extend CT-SGAN to conditional CT-SGAN to avoid external algorithms for the inclusion or exclusion of nodules.% and also explore how the model performs when is provided with a larger number of training samples. Improving small-size nodule detection leaves room for further improvement of the model as well.

\newpage
\bibliographystyle{splncs04}
\bibliography{ct_sgan}

\clearpage
\appendix
\counterwithin{figure}{section} 
\section*{Appendix}
\section{Sample Synthetic CT-scans from CT-SGAN}
%In Figure~\ref{fig:more} we present more samples from our proposed model CT-SGAN.

\begin{figure}[h!] \label{appendix:more1}
	\centerline{\includegraphics[scale=0.62]{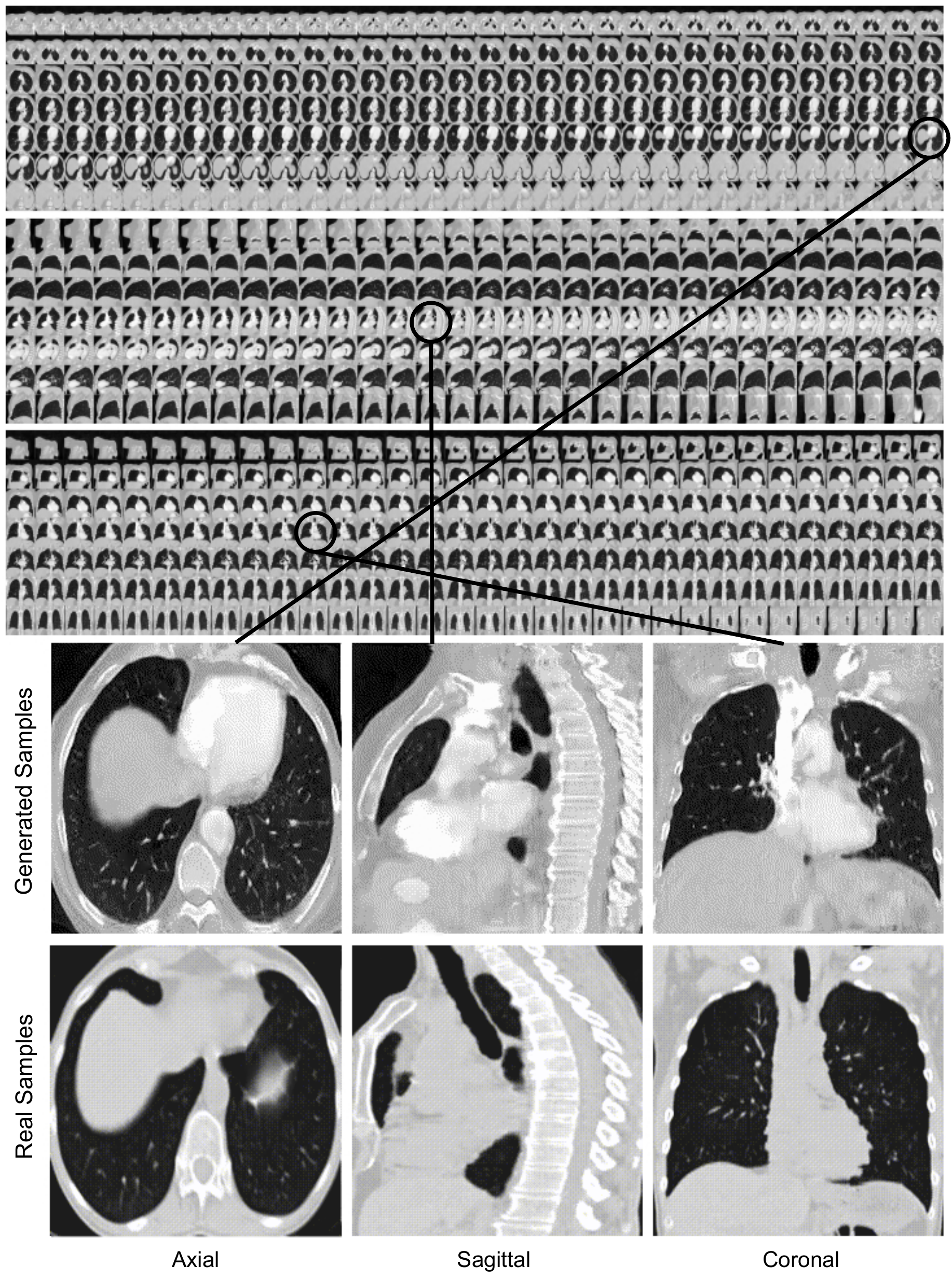}}
	\caption{Middle-region CT-SGAN samples compared with real samples of the same positions.}
\end{figure}

\begin{figure}[h!] \label{appendix:more2}
	\centerline{\includegraphics[scale=0.46]{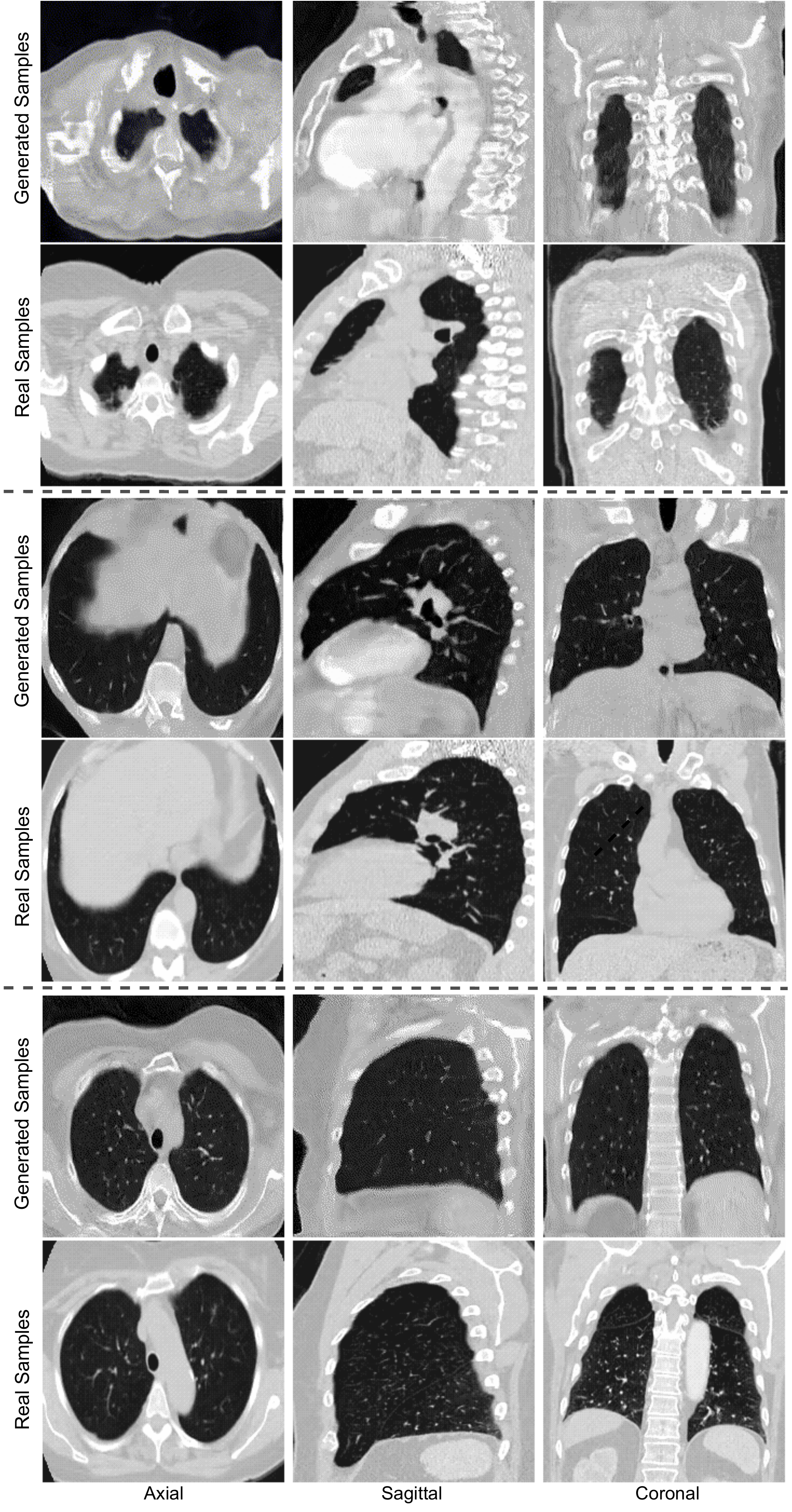}}
	\caption{Comparison between several samples of real and CT-SGAN generated CT, from various corresponding regions. The figure shows how the CT-SGAN generated images, respect the anatomy of the lung.}
\end{figure}

\section{Nodule Injector and Eraser}

\begin{figure}[h!] \label{appendix:injector}
	\centerline{\includegraphics[scale=0.45]{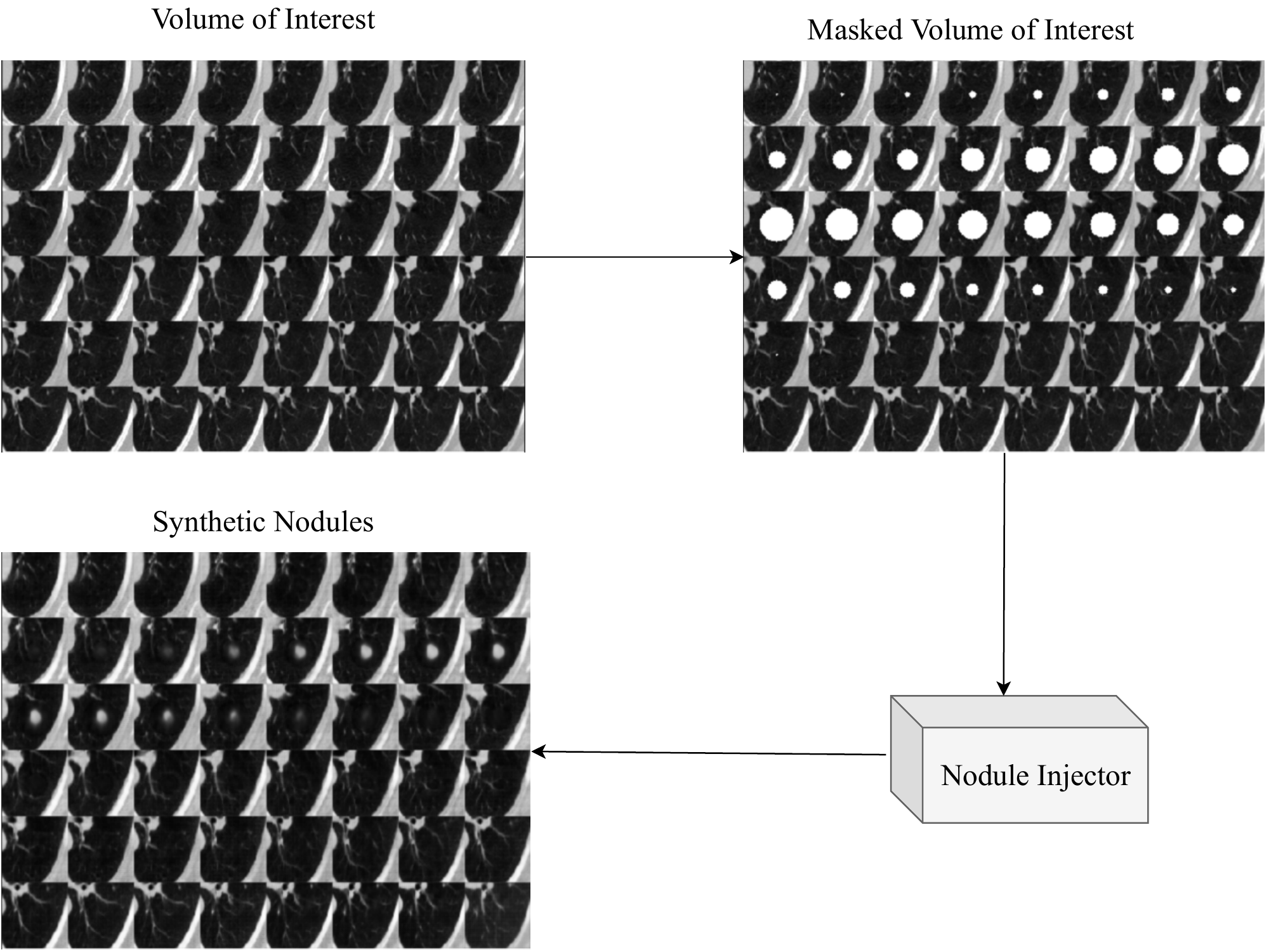}}
	\caption{Illustration of the Nodule Injector during inference mode. Volume of interests (VOIs) were selected as a sub-volume, and a mask was applied to the centre of VOIs as input for the Nodule Injector. The output was the injected synthetic nodules, and the injected VOIs will be pasted back to the CT-scans. }
\end{figure}

\begin{figure}[h!] \label{appendix:eraser}
	\centerline{\includegraphics[scale=0.45]{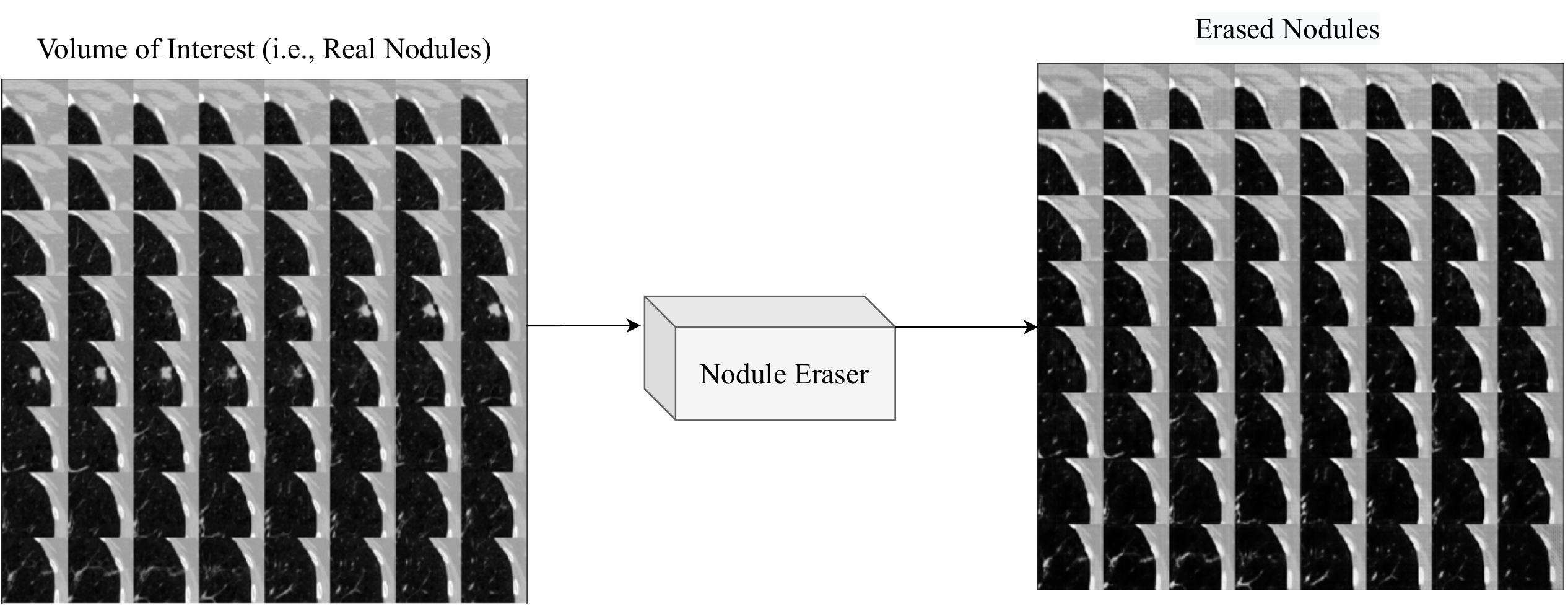}}
	\caption{Illustration of the Nodule Eraser during inference mode. Volume of interests (VOIs) were selected as a sub-volume as input for the Nodule Eraser. The output was the sub-volume without the real nodules, and the erased VOIs will be pasted back to the CT-scans. }
\end{figure}

\begin{figure}[h!] \label{appendix:nodule-radius-dist}
	\centerline{\includegraphics[scale=0.4]{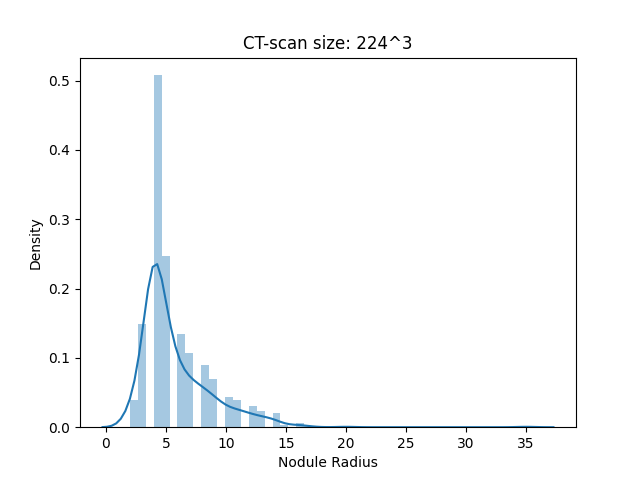}}
	\caption{LIDC Nodule Radius Distribution. }
\end{figure}

\end{document}